# Fabrication of the HFVMTF Double-Bath Cryostat

**R Bruce, V Nikolic, T Tope, O Oshinowo, G Velev**

Fermi National Accelerator Laboratory, Batavia, Illinois 60510, USA

**T Vo, M Maurisak**

Ability Engineering Technology Inc, South Holland, Illinois 60473, USA

Email: rbruce@fnal.gov

**Abstract**. The High Field Vertical Magnet Test Facility (HFVMTF) at Fermilab is designed to test superconducting magnets up to 20 tons and 1.3 meters in diameter. Central to the facility is a double-bath superfluid helium cryostat reaching 1.8 K at 1.2 bar. Integrated with a 15 T superconducting dipole from LBNL, HFVMTF supports HTS cable testing for fusion applications. This paper presents the design and fabrication of the cryostat, with a focus on two complex components: the 2 K heat exchanger and the lambda plate. The heat exchanger thermally links sub-atmospheric and pressurized helium baths. The 1.4-meter lambda plate provides thermal isolation and structural support for over 20 tons under 1.3 bar pressure differential. Finite Element Analysis validated the vessel's integrity under maximum load conditions.

## 1. Introduction

The High Field Vertical Magnet Test Facility (HFVMTF) is a major infrastructure project at Fermilab supporting the DOE Magnet Development Program, with the goal of testing high-field superconducting magnets and High Temperature Superconductor (HTS) samples [1, 2]. Designed to match or exceed the capabilities of leading European test stands such as EDIPO at PSI [3] and FRESCA2 at CERN [4], the facility centers around a large, double-bath cryostat engineered to operate at 1.9 K and 1.2 bar. The cryostat, fabricated by Ability Engineering Technology Inc. in compliance with ASME Boiler and Pressure Vessel Code, will house the superconducting dipole magnet currently under development at Lawrence Berkeley National Laboratory (LBNL) [6].

The helium vessel is supported by a top flange and designed for internal pressures up to 6.9 bar. It can accommodate magnets up to 20 tons in weight, with a maximum diameter of 1.4 m and length of 3 m (see Fig. 1). Its core feature is the 1.4-meter-diameter lambda plate, mounted on a horizontal lambda ring, which provides both thermal separation between the 4.5 K liquid helium and the pressurized superfluid helium baths and structural support for the magnet.

To enhance thermal performance, the design incorporates a ring-shaped saturated superfluid vessel operating at 0.03 bar, located beneath the main helium bath. Thermal coupling between the two baths is achieved using a liquid-to-liquid heat exchanger made of 30 copper U-tubes, a design approach similar to that used at CERN [4]. A compact Joule-Thomson (JT) heat exchanger, manufactured by DATE, is installed at the gas outlet of the saturated vessel to pre-cool the helium before expansion through the JT valve.

A copper thermal shield, cooled by circulating liquid nitrogen, surrounds the helium vessel to reduce radiative heat loads. During transportation, a support pin is temporarily installed at the base of the helium vessel to mitigate mechanical stress when the cryostat is in a horizontal orientation. This pin is removed prior to operation.

This paper presents the final design and fabrication of the HFVMTF cryostat, with particular emphasis on its two most complex components: the lambda plate and the 2 K heat exchanger. The first section reviews the final design, supported by Finite Element Analysis (FEA) of the helium vessel. The second section addresses the main fabrication challenges, including precise alignment and repeated dimensional surveys of the lambda plate during horizontal assembly.

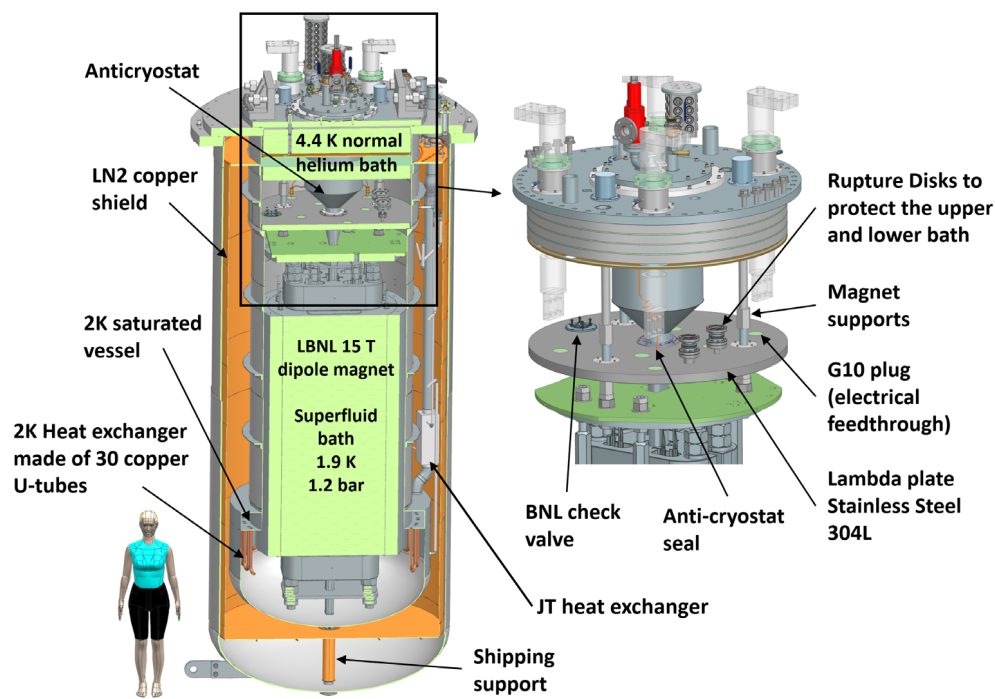

**Figure 1.** Conceptual 3D model of the HFVMTF cryostat (left) and its lambda plate (right)

## 2. Final design of the cryostat
The first section presents the overall design of HFVMTF cryostat.

### 2.1. Helium vessel and internal piping design
As mentioned in the introduction the two most complicated parts of this cryostat are the 2K saturated vessel and the lambda ring (see Fig. 2). The 2K saturated is designed in accordance with the pressure and vessel code and sustain a pressure difference of 6.9 bar during the quench of the superconducting magnets. To achieve this goal the conceptual design, which was copying the CERN design, was modified. Now the vessel has this S-shape which distribute the pressure uniformly at the interface between the superfluid and saturated bath. The copper tubes that thermally connect the two baths, are now brazed on top of 25 mm diameter stainless steel pipes that connect these pipes to the wall of the vessel. In order to sustain the hoop stress at the bottom of the vessel that will elongate the entire vessel and deform the 2K saturated vessel, reinforcement gussets have been placed inside this vessel to reduce the stress on its walls. Due to the complexity of this part and to optimize its design a specific FEA have been performed following the pressure vessel code. This FEA is detailed in the next section.

In addition, the second most complicated part of this design is the lambda ring. This part shall sustain the weight of the magnet as well as the high pressure. Due to the large diameter of this part (1400 mm internal diameter), it has been decided early in the project to use a flat surface 38 mm wide on the cryostat side and an energized seal on the lambda plate side to properly seal the two volumes. Similar methods have been used on BNL cryostat [5, 7]. In this case, the compress® software have been used to define the shape and thickness of this part. The outer diameter of the lambda ring has been reinforced by an outer ring gusset to limit the deformation of this component due to the weight of the magnet and the maximum pressure difference of 1 bar that shall be considered between the upper and lower section of the lambda plate. Here again, due to the complexity of the ring design, a specific FEA shall be performed to validate the design.

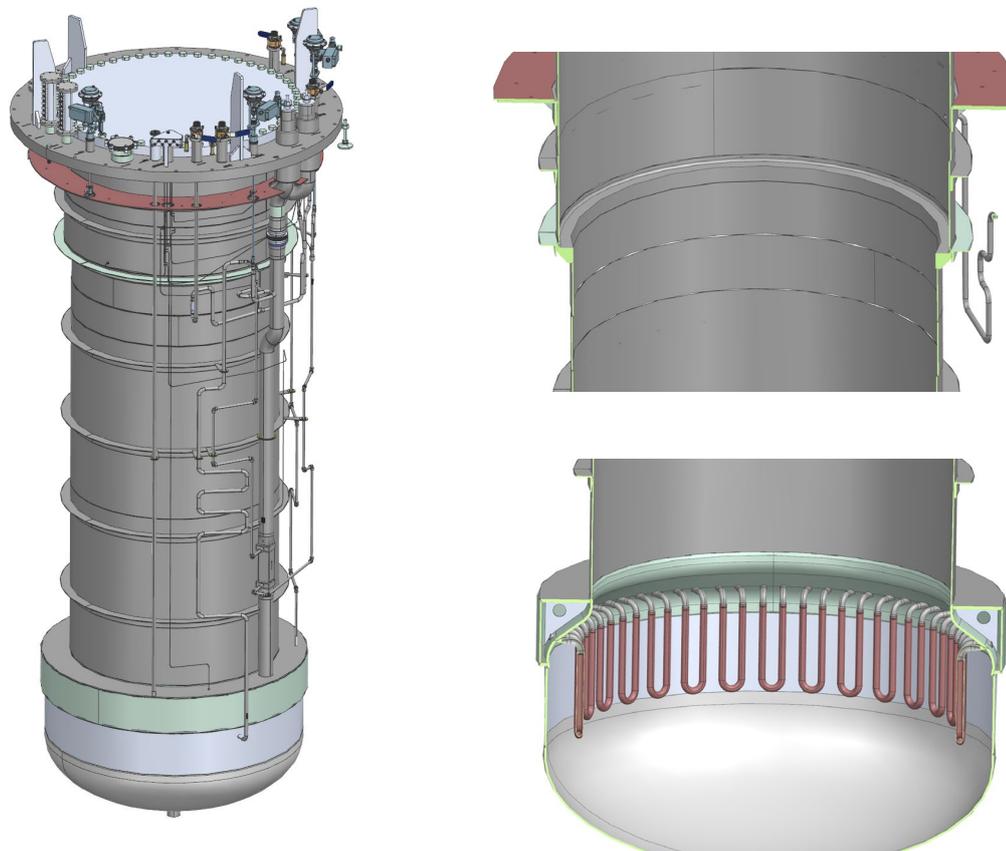

**Figure 2.** 3D model of HFVMTF helium vessels (left), with sectional views of the lambda ring and 2K heat exchanger (right)

*2.2. Helium vessel overall FEA*
Finite Element Analysis (FEA) was conducted by both Fermilab and Ability Engineering to evaluate stress and displacement in the helium vessel under various conditions, including maximum pressure and transport scenarios. This paper presents only Fermilab's results under maximum pressure. A quarter-symmetry model was developed using over 3.6 million nodes and 1.1 million elements, representing the vessel's stainless steel 304L structure. The model incorporated thermal gradients from 300 K at the top flange to 2 K at the lambda ring, a 108,000 lbf load on the lambda ring surface, and internal pressure of 6.9 bar, corresponding to the worst operating case. Initial elastic analysis showed a maximum displacement of 11.86 mm, primarily due to thermal contraction. Stress results, however, exceeded the allowable 138 MPa in localized regions, particularly near the lambda ring and saturated vessel. Stress linearization across thicknesses revealed that the design could not meet ASME BPVC VIII-2 requirements under elastic assumptions.

An elastic-plastic analysis was subsequently performed using material properties and stress-strain curves defined by ASME codes (see Fig. 3). This analysis demonstrated that, despite localized high stresses, the maximum local failure ratio was 0.733—below the critical value of 1—indicating compliance with BPVC standards. The analysis also validated the structural integrity of the vessel under operating thermal conditions and a conservative 3.5 safety factor on pressure loads, confirming no risk of global collapse or local failure.

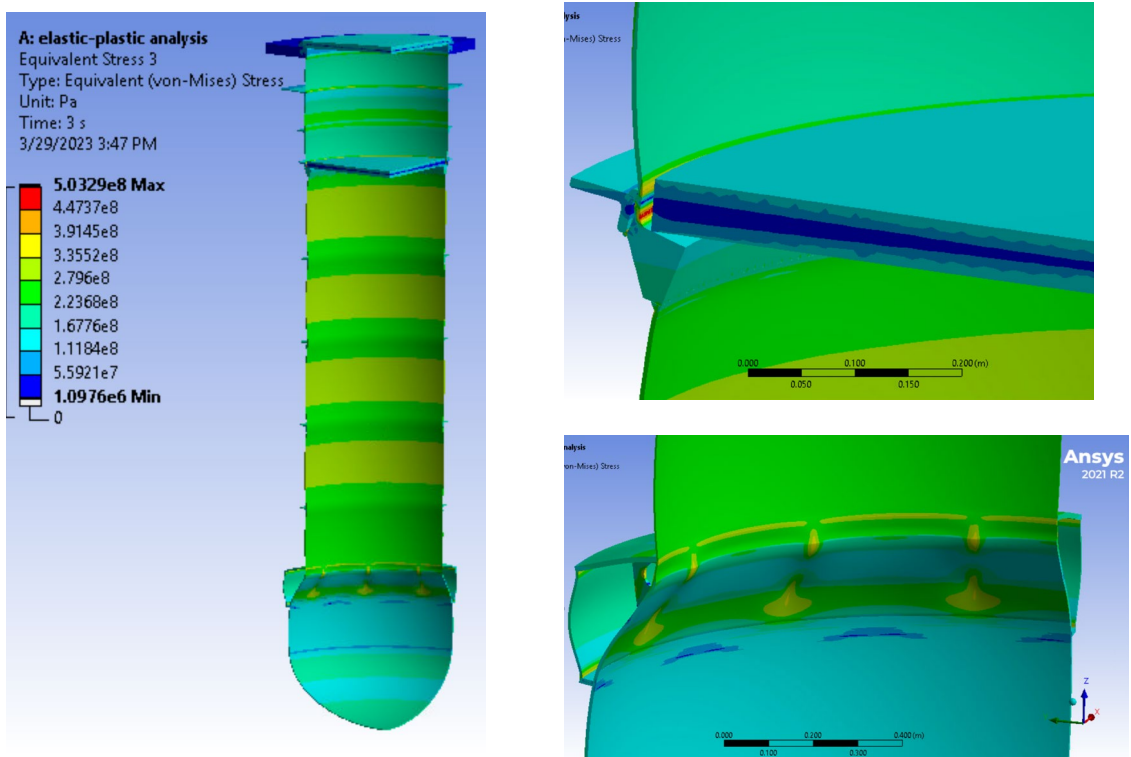

**Figure 3.** Results of the elastic-plastic analysis on the helium vessel

## 2.3. Thermal shield

To limit the heat load to the superfluid bath, G10 supports have been installed between the vacuum vessel and the thermal shield to support it. A thermal FEA have been performed to verify the maximum temperature gradient is lower than 5 K (see Fig. 4). The pressure drop and the evolution of the liquid/gas ratio has been estimated to be less than 140 mbar and 20% of gas volume respectively. For additional safety, a relief pipe has been added to protect the shield tube.

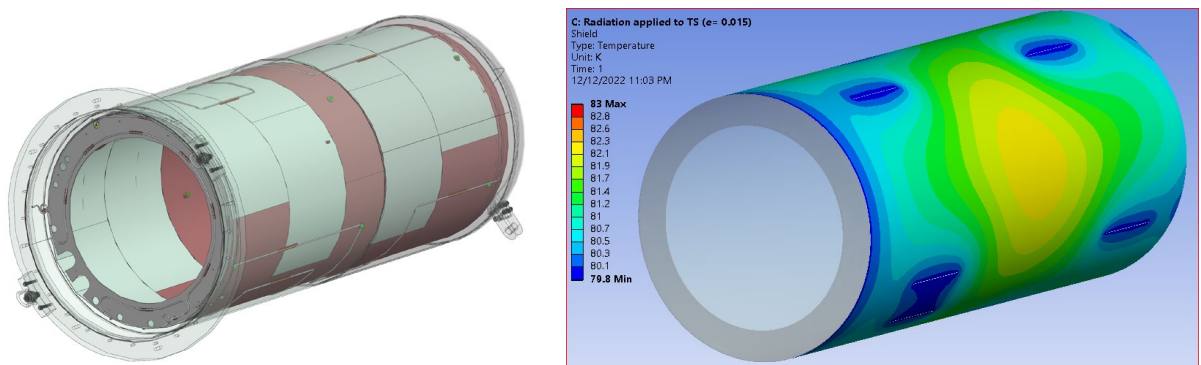

**Figure 4.** Design of the thermal shield on the left and thermal FEA of the shield on the right

In addition, a shipping pin has been designed at the bottom of the cryostat to support the helium vessel during the transportation (see Fig. 5). The maximum acceleration that the cryostat could sustain is 2G horizontal and 5G vertical. As previously mentioned, the cryostat will be shipped horizontal, and it will be move from horizontal to vertical position at Fermilab. When vertical the shipping pin will be removed and replace by a part composed of 5 copper baffles. The closer copper disk to the vacuum vessel will be in contact with the thermal shield using springs, the other baffles will be covered with MLI to limit significantly the radiative heat load to the superfluid bath.

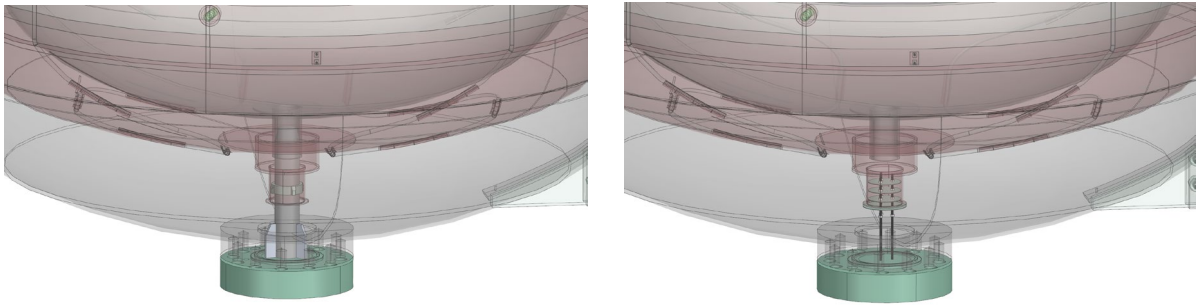

**Figure 5.** Design of the shipping pin on the left and interchangeable part composed of copper baffles on the right

## 3. Fabrication of the cryostat
This section describes the challenges encountered during the fabrication of this cryostat at Ability.

*3.1. Fabrication steps*
The helium vessel was fabricated in several sections and assembled within a massive carbon steel support structure measuring 4 meters in height and approximately 7.6 meters in length (see Fig. 6). This structure is designed to sustain the full weight of the cryostat, roughly 11 tons. The first component fabricated was the 2 K saturated vessel, described in detail in the following section. Next, the top flange—welded to the first cylindrical section of the vessel and incorporating the neck intercept with the upper portion of the copper shield—was mounted onto the support. All cylindrical sections were then carefully aligned within a 3 mm deviation from center and welded together.

Fermilab's survey team visited Ability Engineering to verify the vessel's concentricity and outer diameter. During this visit, they also marked the precise location for the pin holder at the base of the cryostat.

After these verifications and an initial pressure and leak test on the vessel, the control valve bodies, and helium piping were welded to the vessel. Valve openings were inspected, followed by a second round of pressure and leak testing to confirm integrity.

The helium vessel was then wrapped in 15 layers of Multi-Layer Insulation (MLI). Finally, the thermal shield made of copper was slid into place using temporary carbon steel support rings to bear its weight and minimize deformation during lifting. The shield is welded to the neck intercept and wrapped in 45 layers of MLI. The vacuum vessel was then installed using the same method.

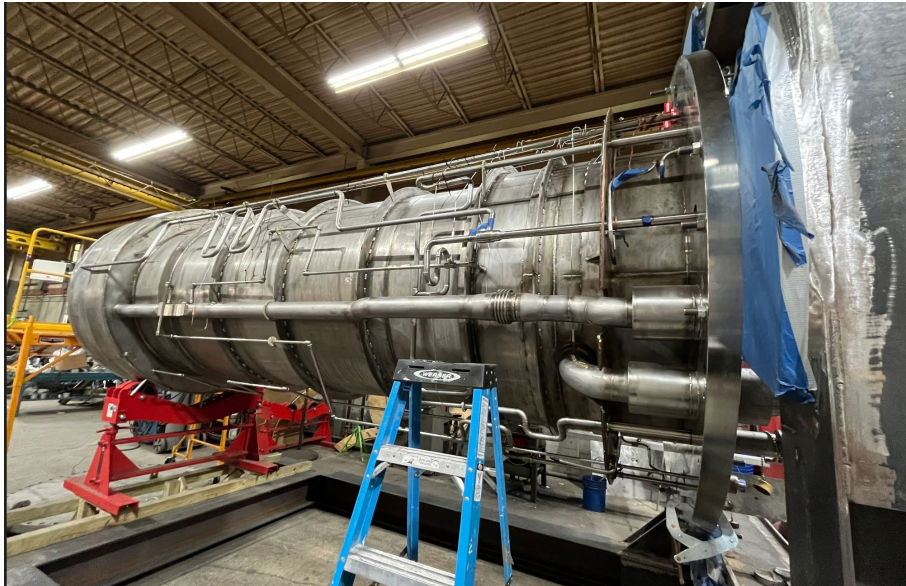
**Figure 6.** Photo of the helium vessel horizontal and attached to its support during fabrication

*3.2. 2K saturated vessel*
The 2 K saturated vessel was challenging to design and fabricate. This cylindrical structure was initially assembled in a vertical orientation, with all copper tubes, internal reinforcement gussets, and end closures completed and pressure-tested (see Fig. 7) before being welded horizontally to the rest of the helium vessel.

The vessel incorporates 30 copper U-tubes, each silver-brazed on top of stainless-steel tubes with a 50 mm overlap. Every joint was leak-tested individually prior to final assembly. Before full-scale fabrication, multiple sample joints were produced and sectioned using Electrical Discharge Machining (EDM) to evaluate the penetration depth of the brazing material between the copper and stainless-steel components.

To protect the brazed joints during the subsequent welding process, a 100 mm gap was maintained between the brazed connections and the vessel wall. After all U-tubes were welded to the vessel and the internal reinforcements installed, the entire assembly was subjected to a pressure and leak test at 6.9 bar to ensure structural integrity and vacuum tightness.

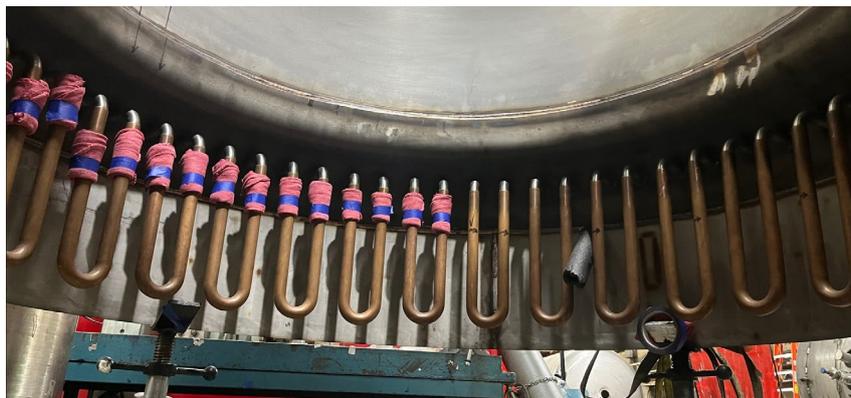
**Figure 7.** Photo of the copper tubes composing the 2K heat exchanger

*3.3. Helium vessel and lambda ring*
The lambda ring was the most complex component to fabricate due to its demanding flatness requirements. To ensure proper sealing of the energized gasket on the lambda plate, the ring's surface had to maintain a flatness within ±0.05 mm over a 1400 mm diameter. This seal is critical, as any leakage between the 2 K superfluid bath and the 4.5 K helium bath would introduce significant heat loads to the 2 K system.

To minimize surface deflection, the lambda ring was CNC machined, and the adjacent upper and lower cylindrical sections of the helium vessel were welded to it. After welding, the ring surface was re-machined using a precision milling process. The welds were intentionally performed at least 500 mm away from the sealing surface to reduce distortion. Despite these precautions, the approach was unsuccessful. A survey conducted by Fermilab using a laser tracker revealed a flatness deviation of ±1 mm (±0.04" in Fig. 8), exceeding the allowable tolerance by a factor of 20.

Fermilab and Ability Engineering decided to correct the surface after welding of the entire helium vessel to avoid future issues. The assembly was sent to a subcontractor specializing in large rotating equipment. A 6-foot-long boring bar was mounted to the top flange with internal supports positioned below the lambda ring (see Fig. 8). The ring surface was surveyed before and after installation of the boring bar, and again following final machining. The final survey confirmed the flatness was brought within the required tolerance (see Fig. 9). After this machining, the lambda ring surface was polished to achieve a roughness average lower than 0.2 µm, also required by the seal manufacturer.

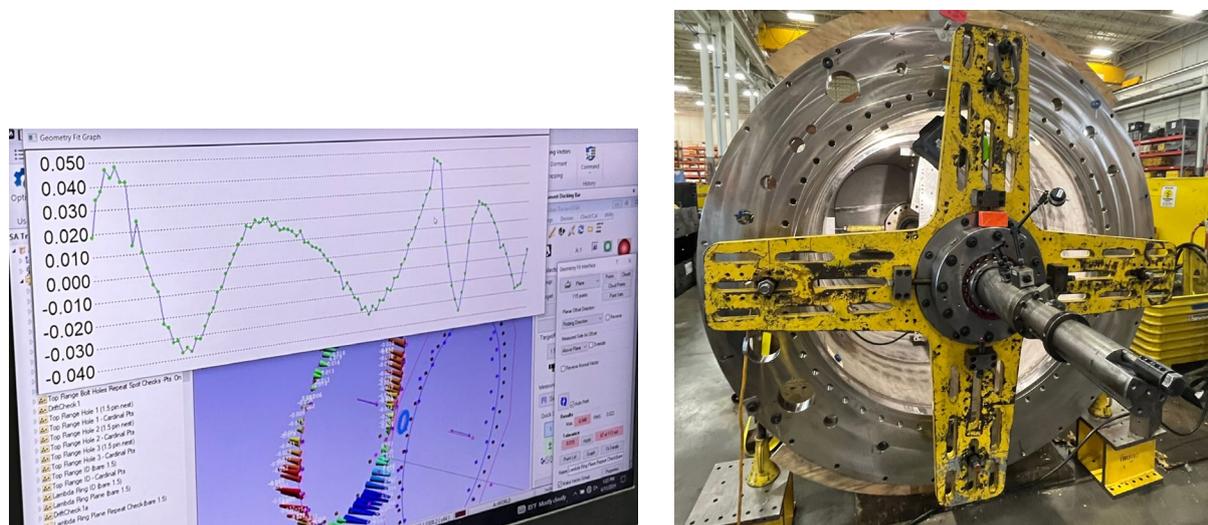

**Figure 8.** First Fermilab's survey for the lambda ring on the left (units in inches) and rectification of the lambda ring surface of the right

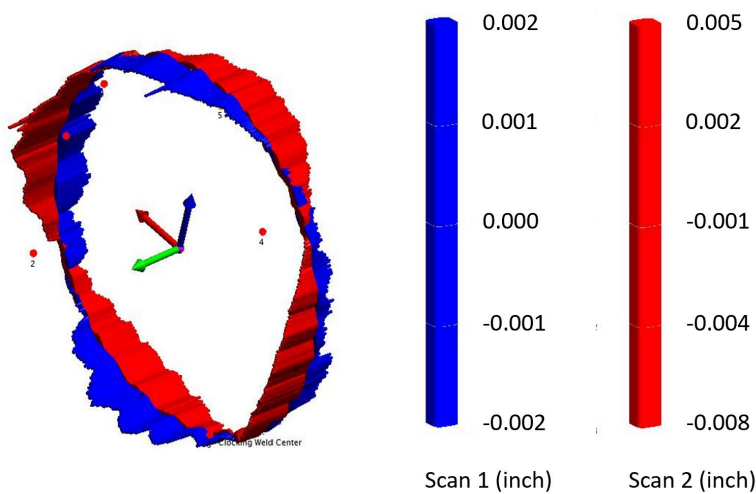

**Figure 9.** Second Fermilab's survey for the lambda ring

## 4. Conclusion

Designed to accommodate superconducting magnets up to 20 tons in weight and operating at 1.9 K, the HFVMTF cryostat incorporates a range of technical innovations—including a high-precision lambda plate interface, a dual-bath architecture, and a custom-engineered 2 K heat exchanger—to meet the stringent thermal and mechanical requirements of high energy superconducting magnets.

The design process was guided by rigorous analysis and adherence to ASME BPVC standards, including extensive finite element modelling under various load cases. Both elastic and elastic–plastic analyses confirmed the mechanical integrity of the helium vessel and validated compliance with pressure vessel code requirements.

The fabrication phase presented several challenges, particularly the integration of the 2 K saturated vessel and the machining of the lambda ring to the required flatness. These challenges were successfully addressed through close collaboration between Fermilab and industry partners, advanced machining techniques, and detailed metrology at each critical step.

## 5. References

[1] Velev G V et al 2021 Design and construction of a high field cable test facility at Fermilab *IEEE Trans. on Appl. Supercond* vol 31 pp 1–4
[2] Velev G V et al 2022 Status of the High Field Cable Test Facility at Fermilab *IEEE Trans. on Appl. Supercond*
[3] Bruzzone P et al 2016 EDIPO: The Test Facility for High-Current High Field HTS Superconductors *IEEE Trans. Appl. Supercond.* vol 26 no 2 pp 35-40
[4] Verweij A P et al 1999 1.9 K test facility for the reception of the superconducting cables for the LHC *IEEE Trans. Appl. Supercond.* vol 9 no 2 pp 153-156
[5] Koshelev S et al 2022 Design of the cryostat for High Field Vertical Magnet Testing Facility at Fermilab *IOP Conf. Ser.: Mater. Sci. Eng.* 1240 012081
[6] Vallone G et al 2021 Magnetic and Mechanical Analysis of a Large Aperture 15 T Cable Test Facility Dipole Magnet *IEEE Trans. on Appl. Supercond.* vol 31 pp 1–6.
[7] Bruce R et al 2023 Cryogenic and safety design of the future high field cable test facility at Fermilab *IOP Conf. Ser.: Mater. Sci. Eng.* 1301 012093

**Acknowledgments**

This manuscript has been authored by FermiForward Discovery Group, LLC under Contract No. 89243024CSC000002 with the U.S. Department of Energy, Office of Science, Office of High Energy Physics. FERMILAB-CONF-25-0328-TD